\documentclass[aps,amsmath,nofootinbib,superscriptaddress,amssymb,preprintnumbers,floatfix,showpacs,preprint]{revtex4-1}
 \usepackage{amsmath,txfonts,longtable,booktabs,overpic,amssymb,bm,bbm,multirow,float,graphicx,color,dcolumn,subfigure,hyperref,tikz}
\definecolor{blue}{RGB}{45,48,146}
 \hypersetup{colorlinks,citecolor=black,anchorcolor=black,menucolor=black, linkcolor=black,filecolor=black,runcolor=black,urlcolor=black,frenchlinks=true}

\begin{document}

	\title{Vortex structures in electron-positron pair production by two-colored fields}

  \author{Adiljan Sawut}
	\affiliation{ State Key Laboratory for Tunnel Engineering, China University of Mining and Technology, Beijing 100083, China}
	
	\author{Ying-Jun Li}\email{lyj@aphy.iphy.ac.cn}
	\affiliation{ State Key Laboratory for Tunnel Engineering, China University of Mining and Technology, Beijing 100083, China}

\author{Hong-Hao Fan}
	\affiliation{Key Laboratory of Beam Technology of the Ministry of Education, and School of Physics and Astronomy, Beijing Normal University, Beijing 100875, China}

 \author{Bai-Song Xie}\email{bsxie@bnu.edu.cn}
	\affiliation{Key Laboratory of Beam Technology of the Ministry of Education, and School of Physics and Astronomy, Beijing Normal University, Beijing 100875, China}
\date{\today}

\begin{abstract}
   We investigate the spin-resolved vortex properties of electron–positron pairs created from vacuum in time-delayed, two-color electromagnetic fields. By treating the temporal delay $G$ as a continuous tuning parameter, we reveal a dynamic transition from interference-dominated domain patterns at $G=0$ to the nucleation of quantized vortex lattices at $G=0.5$. These topological structures exhibit a staggered arrangement analogous to von Kármán vortex streets in fluid dynamics. We demonstrate that the momentum-space morphology is strictly governed by spin-orbit selection rules: parallel spin configurations ($\uparrow\uparrow, \downarrow\downarrow$) enforce a dipole-like connectivity, while anti-parallel configurations ($\uparrow\downarrow, \downarrow\uparrow$) resolve into distinct quadrupole structures. This difference originates from the conservation of total angular momentum $J_z$, where the spin projection determines the required orbital angular momentum $L_z$ of the created pairs. At large delays ($G \ge 1$), macroscopic vortex coherence dissolves into a chaotic phase landscape due to multi-channel interference, yet the spin-dependent nodal geometries remain robust. Our findings suggest that these topological signatures provide a high-fidelity diagnostic for the quantum dynamics of vacuum excitations in strong-field QED.
\end{abstract}
\maketitle

\section{Introduction\label{sec:1}}
Nonlinear quantum electrodynamics (QED) in ultrastrong electromagnetic fields has attracted extensive and growing interest, driven by the rapid development of multi-petawatt and exawatt laser facilities. One of the most fundamental predictions of QED is that the vacuum becomes unstable in sufficiently strong external fields, enabling the spontaneous creation of electron–positron pairs—the Sauter–Schwinger effect\cite{Sauter:1931zz,Heisenberg:1936nmg,Schwinger:1951nm,PhysRevD.44.1825,Fradkin:1991,DiPiazza:2011tq,Xie:2017xoj}. Considerable recent progress has been made in understanding the microscopic properties of these created particles, including spin-resolved momentum spectra \cite{Hu2023,Amat:2024nvg,Aleksandrov:2024cqh,Hu:2024nyp,GiesTorgrimsson2016,Otto2015,Chen:2025xib,Kohlfurst:2018kxg,Kohlfuerst2013}, quantum entanglement \cite{Li:2016zyv}, and topological structures such as spirals and phase vortices in momentum space \cite{Fan:2024nsl,Li:2017qwd}.

Phase vortices are universal topological features arising in many fields of physics \cite{berry,PMDirac:1931,BialynickiBirula1992TheoryOQ,Bialynicki-Birula:2016unl}, from molecular vibrations \cite{Luski:2021} and electron vortex beams \cite{Ivanov:2022jzh,Bliokh:2017} to hydrodynamic flows \cite{Bliokh:2025} and laser-driven quantum dynamics \cite{Ngoko:2015,Hebenstreit:2016xhn,Pengel:2017,Majczak:2022xlv,Geng2020VortexSI}. In strong-field physics, optical OAM beams and electron vortices have been generated and controlled experimentally, revealing rich angular-momentum transfer mechanisms between fields and particles \cite{Sukhorukov:2005,Bliokh:2012,Jhajj:2016,Hancock:2019,Huang:2022}. Notably, vortex–antivortex structures have also been identified in the complex probability amplitude of strong-field ionization \cite{Pengel:2017,Majczak:2022xlv,Cajiao:2020} and in vacuum pair production under circularly polarized fields \cite{Bechler:20232024,Majczak:2024}. These observations highlight the importance of vortex topology for understanding the fundamental dynamics of QED in strong fields.

More recently, vortex features have also been predicted in vacuum pair production, particularly within the multiphoton regime \cite{Fan:2024nsl}. For scalar QED, it has been shown that the topology of momentum-space vortices reflects the intrinsic OAM of the produced scalar particles, and is strongly correlated with the number and helicity of absorbed photons \cite{Bliokh:2022l,Bliokh:2011fi}. However, scalar theories ignore the rich spin structure that naturally arises in the fermionic case. In spinor QED, the total angular momentum transferred to electron–positron pairs contains both intrinsic spin and orbital components \cite{Bliokh:2022l}. Systematic understanding of how the spin configuration influences vortex formation, evolution, and annihilation in strong fields remains largely absent.

In addition, most previous works utilize single-frequency circularly polarized \cite{Geng2020VortexSI} or linearly polarized laser fields \cite{Cajiao:2020}. The behavior of vortex structures under two-color fields with tunable time delay, which allow precise control over the temporal angular-momentum transfer process, has not yet been explored in vacuum pair production. The relative phase between the two components introduces a new degree of freedom enabling the controlled birth, deformation, and disappearance of vortices, making such configurations ideal for manipulating topological features of the created particles.

In this work, we investigate the spin-resolved vortex properties of electron--positron pairs generated by two-color electric fields with a controllable time delay $G$. By employing the Bogoliubov transformation framework, we analyze how the momentum-space topology depends on the temporal overlap of the field components. We show that the total topological charge is governed by spin-orbit selection rules and the conservation of total angular momentum $J_z$. Specifically, we identify a topological transition from trivial phase domains to organized lattices analogous to von Kármán vortex streets in fluid dynamics \cite{Cajiao:2020,Acheson1990}. Our results demonstrate that while macroscopic vortex structures are sensitive to multi-channel interference at large delays, the spin-dependent nodal geometries (dipole vs. quadrupole) remains a robust signature of the vacuum excitation.

We use the atomic units ($\hbar=m_e=e=1$) throughout this paper.

Our paper is organized as follows. In Sec.~\ref{sec:2}, we introduce the external field and the particle distribution function, followed by a review of the vortex description. In Sec.~\ref{sec:3}, we present the numerical results. The Sec.~\ref{sec:4} is a summary for our work.

\section{Theoretical method}\label{sec:2}

The time-dependent creation and annihilation operators for an electron with spin $s$ and a positron with spin $s'$, denoted by $\tilde{a}_{p,ss'}(t)$ and $\tilde{b}_{-p,ss'}^\dag(t)$, respectively, are related to their initial operators ${a}_{p,ss'}$ and ${b}_{-p,ss'}^\dag$ through Bogoliubov transformations. The corresponding transformation coefficients $\alpha_{p,ss'}(t)$ and $\beta_{p,ss'}(t)$ satisfy
\begin{equation}
\begin{bmatrix}
\tilde{a}_{p,ss'}(t) \\
\tilde{b}_{-p,ss'}^\dag(t)
\end{bmatrix}
=
\begin{bmatrix}
\alpha_{p,ss'}(t) & \beta_{p,ss'}^*(t) \\
\beta_{p,ss'}(t) & \alpha_{p,ss'}^*(t)
\end{bmatrix}
\begin{bmatrix}
a_{p,ss'} \\
b_{-p,ss'}^\dag
\end{bmatrix}.
\end{equation}
Here, $|\beta_{p,ss'}(t)|^2$ represents the momentum distribution of the produced particles with canonical momentum $p$. Due to fermionic statistics, these coefficients must obey the constraint
\begin{equation}
\sum_{ss'}\left(|\alpha_{p,ss'}(t)|^2 + |\beta_{p,ss'}(t)|^2\right) = 1.
\end{equation}

The evolution of the Bogoliubov coefficients is governed by
\begin{equation}
\frac{d\alpha_{p,ss'}(t)}{dt} = \Omega_{p,ss'}(t) \beta_{p,ss'}(t) \, e^{2i\int^{t} d\tau\, \omega_p(\tau)},
\end{equation}
\begin{equation}
\frac{d\beta_{p,ss'}(t)}{dt} = \Omega_{p,ss'}(t) \alpha_{p,ss'}(t) \, e^{2i\int^{t} d\tau\, \omega_p(\tau)},
\end{equation}
where $\Omega_{p,ss'}(t)=u_{p,s}^\dagger(t)\dot{H}_{p}(t)v_{p,s'}(t) / [2\omega_p(t)]$. The spinors $u_{p,s}(t)$ and $v_{p,s'}(t)$ are instantaneous eigenstates of the Dirac Hamiltonian
\begin{equation}
H(t)=\boldsymbol{\alpha}\cdot(\mathbf{p}-e\mathbf{A}(t)) + \beta m,
\end{equation}
and the corresponding energy reads
\begin{equation}
\omega_p(t) = \sqrt{m^2 + q^2(t)} = \sqrt{m^2 + [p - eA(t)]^2},
\end{equation}
with kinetic momentum $q(t)=p-eA(t)$.

For convenience, we introduce the phase-rotated coefficients
\begin{equation}
c^\alpha_{p,ss'}(t) = e^{-i\int d\tau\,\omega_p(\tau)}\alpha_{p,ss'}(t), \qquad
c^\beta_{p,ss'}(t) = e^{i\int d\tau\,\omega_p(\tau)}\beta_{p,ss'}(t),
\end{equation}
which lead to a compact two-level dynamical form:
\begin{equation}
i\frac{d}{dt}
\begin{bmatrix}
c^\alpha_{p,ss'}(t) \\
c^\beta_{p,ss'}(t)
\end{bmatrix}
=
\begin{bmatrix}
\omega_p(t) & -i\Omega_{p,ss'}(t) \\
-i\Omega_{p,ss'}(t) & -\omega_p(t)
\end{bmatrix}
\begin{bmatrix}
c^\alpha_{p,ss'}(t) \\
c^\beta_{p,ss'}(t)
\end{bmatrix}.
\end{equation}

Throughout this work, the spin refers to the projection of the electron–positron pair along the $z$-axis, defined as $S_z=s_z+s'_z$, rather than categorizing the system into singlet or triplet states. The momentum-space distributions are then given by
\begin{equation}
f_{ss'}(p)=2\,|c^\beta_{p,ss'}(t)|^2,
\end{equation}
and the total particle yield reads
\begin{equation}
f(p)=\sum_{ss'} f_{ss'}(p).
\end{equation}

In three-dimensional momentum space, vortex lines are formed as closed loops or extended continuous structures surrounding nodal surfaces of vanishing probability, across which the phase of the probability amplitude exhibits discontinuous jumps of $\pm\pi$. In two-dimensional slices, these vortices appear as isolated points, whereas nodal surfaces reduce to zero-amplitude curves.

The complex amplitude $c^\beta_{p,ss'}(t)$ encodes the complete topological information. The associated spin-resolved Berry connection is defined as
\begin{equation}
\boldsymbol{\mathcal{A}}_{ss'} =
\frac{\operatorname{Re}\left[(c^\beta_{p,ss'})^* (-i\nabla_p)c^\beta_{p,ss'}\right]}
{|c^\beta_{p,ss'}|^2}
=
\nabla_p\left[\arg(c^\beta_{p,ss'})\right],
\end{equation}
which leads to the quantization condition
\begin{equation}
\frac{1}{2\pi}\oint_C \boldsymbol{\mathcal{A}}_{ss'}\cdot d\mathbf{p} = l_{ss'},
\end{equation}
where $l_{ss'}\in\mathbb{Z}$ denotes the spin-dependent winding number, or equivalently, the topological charge. This formalism reveals the intrinsic connection between the phase structure of the pair-production amplitude and its spin-resolved topological properties.

\section{Numerical results}\label{sec:3}
\begin{figure}[ht!]
  \includegraphics[width=0.9\textwidth]{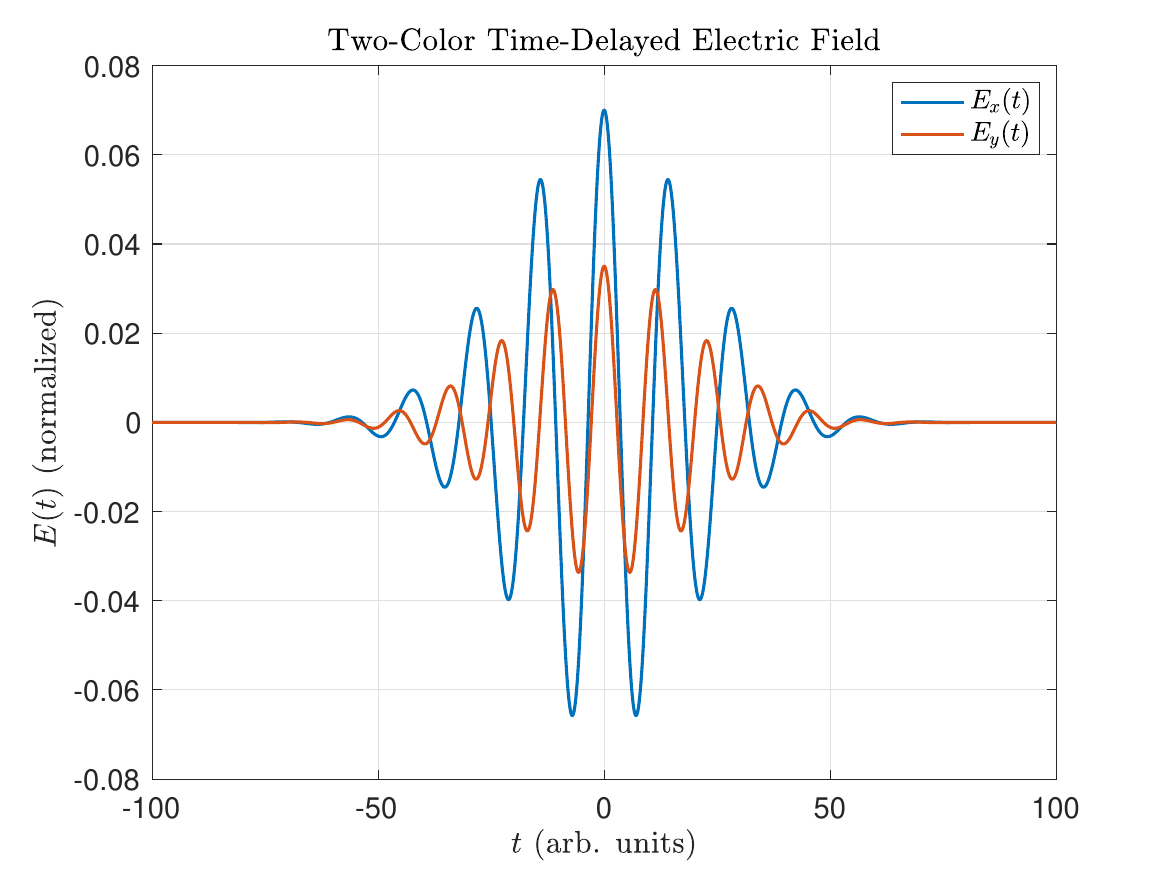}
  \caption{Shape of the two-color fields\label{fig:1}, where the field strengths are $E_x = 0.07E_{\mathrm{cr}}$ and $E_y = 0.035E_{\mathrm{cr}}$, the field frequencies are $\omega_1 = 0.44\,c^2$ and $\omega_2 = 0.55\,c^2$.}
  \end{figure}

In this work, our field configuration is
\begin{equation}
\mathbf{E}(t) =
E_x \exp(-t^2/2\tau_1^{2})
\cos(\omega_1 t)\mathbf{e_x}
+
E_y \exp(-(t - T_d)^2/2\tau_2^{2})
\cos(\omega_2 (t - T_d))\mathbf{e_y},
\end{equation}
where $E_x=0.07E_{\mathrm{cr}}, E_y=0.035E_{\mathrm{cr}}$ gives the external field amplitude in units of the critical field $E_{cr}=m^2/e\approx1.3\times10^{16}\rm{V/cm}$, we set the fields frequencies as $\omega_1=0.44c^2, \omega_2=0.55c^2$ and $\tau_1=\tau_2=20, T_d=G(\tau_1 + \tau_2),$ pulse durations expressed in number of field cycles $N_1=\omega_1 \tau_1\approx9, N_2= \omega_2 \tau_2=11$. The polarization vectors $\mathbf{e_x}$ and $\mathbf{e_y}$ align with the $x-$ and $y-$ directions, as shown in the Fig.~\ref{fig:1}, where the time-delayed parameter $G=0$.

To observe the formation of effects in multiphoton process, we set other parameters unchanged, and vary the time-delay parameter $G$ from 0 to 2 in our entire calculation. In this case, We observe that the vortex topology is highly sensitive to the time delay: strictly absent at $G=0$, it nucleates at small $G$ values before diminishing and eventually vanishing entirely at $G=2$. Most interestingly, the disappearance of vortex structures at $G=0.5$ is attributed to the mutual cancellation of oppositely rotating vortices, mirroring the annihilation dynamics of von Kármán vortex streets \cite{Cajiao:2020}.

  \begin{figure*}[!htb]
      \includegraphics[width=0.24\linewidth]{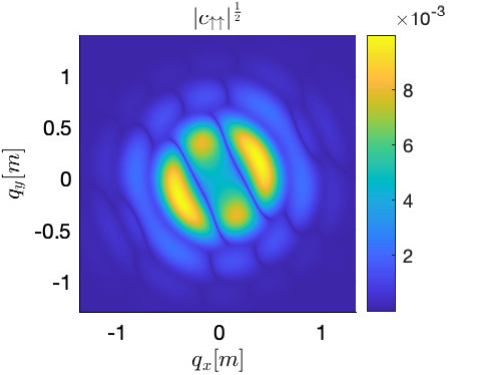}
      \includegraphics[width=0.24\linewidth]{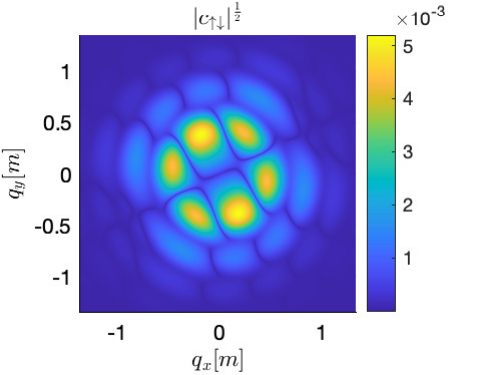}
      \includegraphics[width=0.24\linewidth]{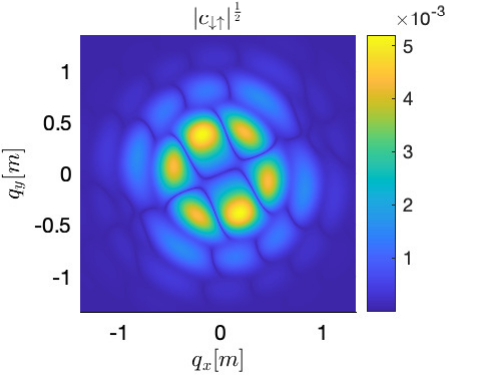}
      \includegraphics[width=0.25\linewidth]{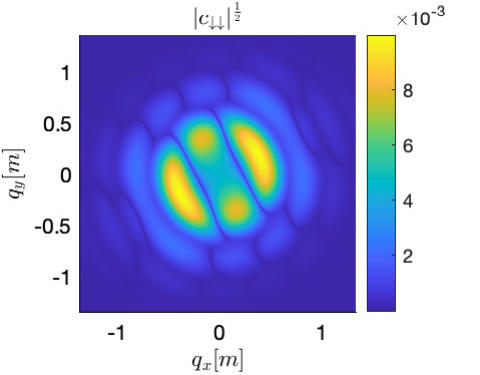}
      \includegraphics[width=0.24\linewidth]{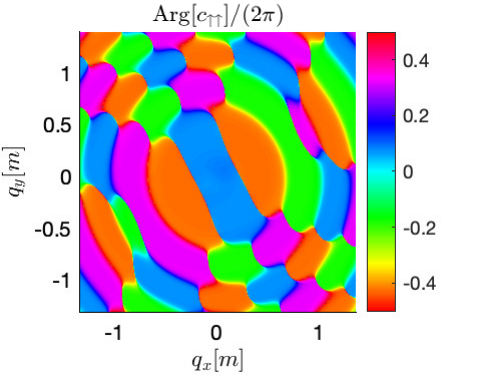}
      \includegraphics[width=0.24\linewidth]{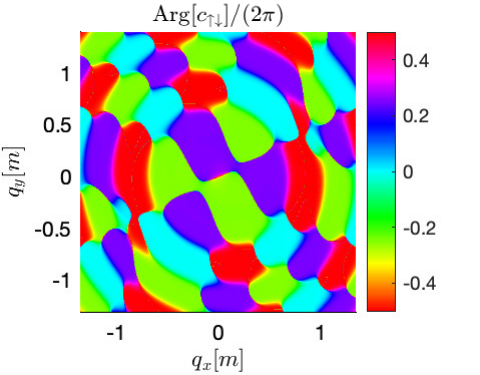}
      \includegraphics[width=0.24\linewidth]{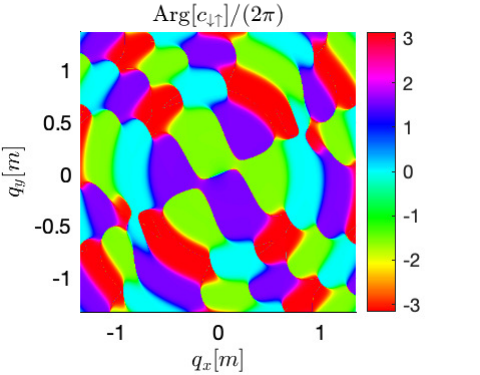}
      \includegraphics[width=0.25\linewidth]{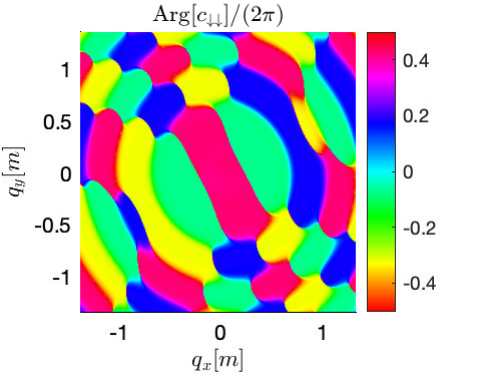}
      \caption{Particle momentum distributions (upper panels) and corresponding phase distributions (lower panels) for electron--positron pair production in the two-color cross field with time-delay parameter $G=0$. From left to right, the spin configurations are $(\uparrow\uparrow)$, $(\uparrow\downarrow)$, $(\downarrow\uparrow)$, and $(\downarrow\downarrow)$. The field strengths are $E_x = 0.07E_{\mathrm{cr}}$ and $E_y = 0.035E_{\mathrm{cr}}$, the field frequencies are $\omega_1 = 0.44\,c^2$ and $\omega_2 = 0.55\,c^2$. The pulse durations are $\tau_1 = \tau_2 = 20$, corresponding to $N_1 = \omega_1 \tau_1 \approx 9$ and $N_2 = \omega_2 \tau_2 = 11$ cycles.
}\label{Fig:Data1}
 \end{figure*}

The absence of vortex structures at the zero-delay limit ($G=0$) is attributed to the high temporal symmetry of the overlapping fields. In this case, the electric field vector $\mathbf{E}(t)$ traces a closed Lissajous-like curve \cite{Lissajous:2020}. Because the frequencies $\omega_1$ and $\omega_2$ are commensurate, the system maintains a temporal parity that suppresses the accumulation of the non-trivial geometric phase (Berry phase) required to produce isolated phase singularities. Consequently, the phase distribution is dominated by linear nodal lines rather than quantized vortices. The upper panels of Fig.~\ref{Fig:Data1} display the particle momentum distributions for four distinct spin configurations for electron-positron pair production, while the lower panels depict the corresponding phase distributions. From left to right, these correspond to the up-up ($\uparrow\uparrow$), up-down ($\uparrow\downarrow$), down-up ($\downarrow\uparrow$), and down-down ($\downarrow\downarrow$) channels. It is shown that the particle distributions are identical for the group up-up, down-down and the group up-down, down-up configurations, while the phase distributions are shifted the color. Furthermore, we observe that the distributions exhibit distinct separation patterns for different spin configurations. In the up–up and down–down cases, the particle distributions split into four primary lobes. In contrast, for the up–down and down–up configurations, the distributions divide into six main lobes concentrated in the central region, as clearly shown in the figure. The distinct separation patterns originate from the spin-dependent structure of the production amplitude. For $G=0$, the momentum distribution is governed by interference among multiple quantum trajectories, whose relative weights are controlled by spin-dependent prefactors. Parallel spin configurations up-up ($\uparrow\uparrow$) and down-down ($\downarrow\downarrow$) preserve a higher degree of symmetry, leading to fewer constructive interference channels and resulting in four dominant lobes in momentum space. In contrast, anti-parallel configurations up-down ($\uparrow\downarrow$), down-up ($\downarrow\uparrow$), reduce this symmetry through relative phase shifts between spin components, thereby activating additional interference channels and producing six pronounced central lobes. This effect is most prominent in the low-momentum region, where contributions from different trajectories strongly overlap.

The phase distributions in the second row do not exhibit vortex structures for Fig.~\ref{Fig:Data1}. In particular, no isolated amplitude zeros accompanied by a $\pm 2\pi$ phase winding are observed. Instead, the phase forms extended domains separated by nodal lines, reflecting interference between different quantum trajectories. The boundaries between these domains correspond to destructive interference in the magnitude distribution shown in the first row. For the parallel spin configurations, the phase pattern remains relatively regular and symmetric, whereas the anti-parallel configurations display a more intricate arrangement of phase sectors, consistent with the enhanced interference responsible for the six-lobe structure in the particle density. Thus, at $G=0$, the phase structure is governed by symmetry-controlled interference rather than by topological vortex excitations.

  \begin{figure*}[!htb]
      \includegraphics[width=0.24\linewidth]{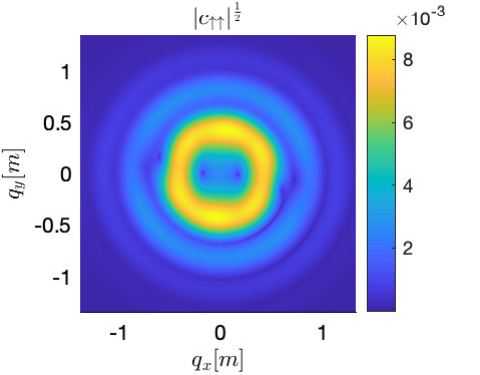}
      \includegraphics[width=0.24\linewidth]{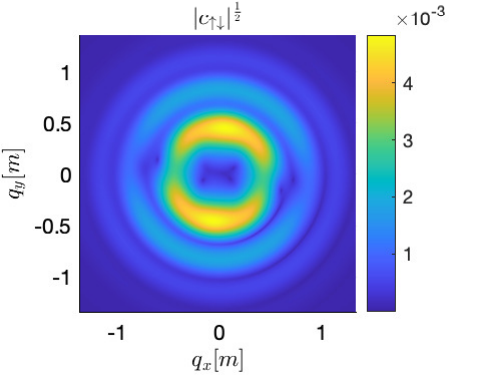}
      \includegraphics[width=0.24\linewidth]{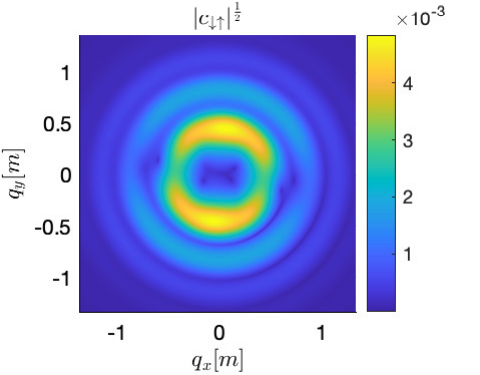}
      \includegraphics[width=0.25\linewidth]{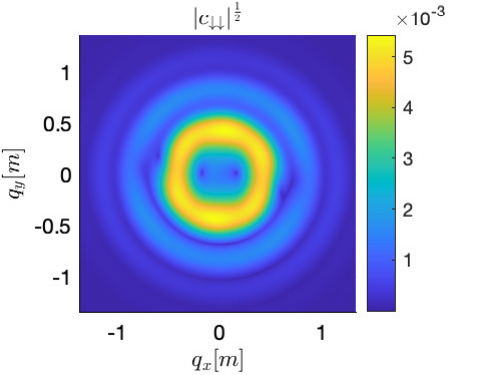}
      \includegraphics[width=0.24\linewidth]{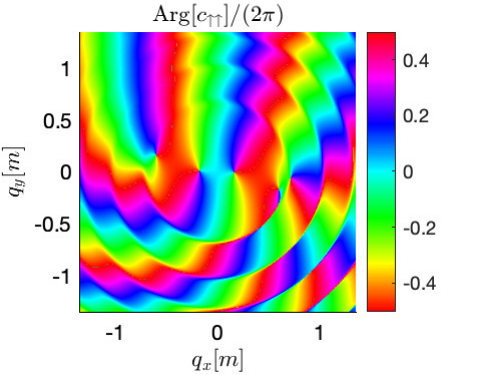}
      \includegraphics[width=0.24\linewidth]{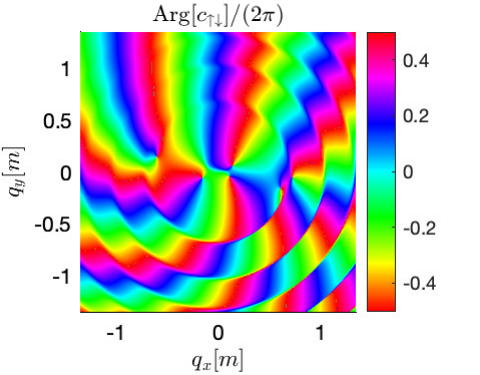}
      \includegraphics[width=0.24\linewidth]{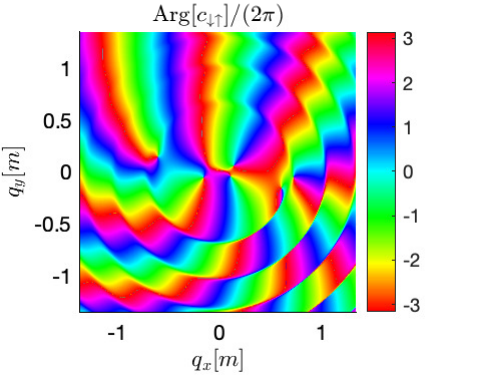}
      \includegraphics[width=0.25\linewidth]{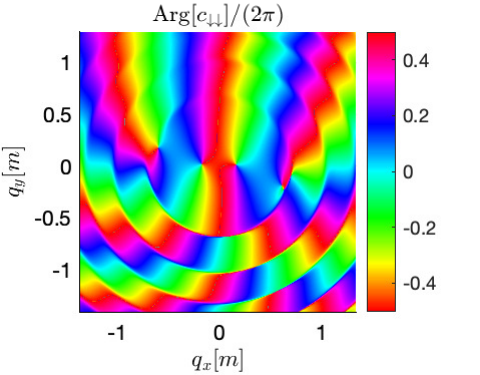}
      \caption{Particle momentum distributions (upper panels) and corresponding phase distributions (lower panels) for electron--positron pair production in the two-color cross field with time-delay parameter $G=0.2$. Other parameters are the same with Fig.~\ref{Fig:Data1}.}\label{Fig:Data2}
 \end{figure*}

 In contrast to the $G=0$ limit shown in Fig.~\ref{Fig:Data1}, the introduction of a small temporal delay $G=0.2$ in  Fig.~\ref{Fig:Data2} initiates a qualitative topological transition from interference-dominated domain patterns to nontrivial vortex configurations. Specifically, the phase distributions exhibit clear vortex structures in momentum space, characterized by isolated amplitude minima near the central region. Around these points, the phase displays continuous spiral winding, accumulating a net $\pm 2\pi$ variation along closed encircling contours to indicate well-defined topological charges. Crucially, these vortex cores coincide with pronounced depressions in the particle distributions (upper panels), confirming their origin as phase singularities of the production amplitude. In this weak-delay regime, the finite $G$ parameter modifies the interference conditions and breaks the symmetry that previously suppressed topological excitations.

 \begin{figure*}[!htb]
      \includegraphics[width=0.24\linewidth]{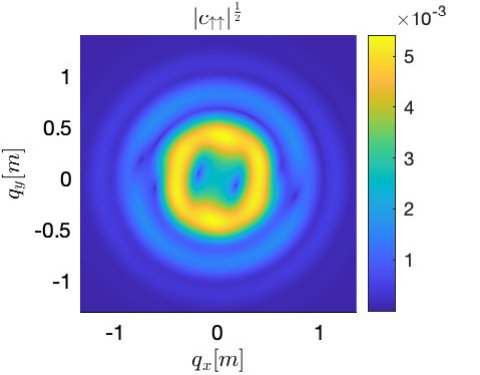}
      \includegraphics[width=0.24\linewidth]{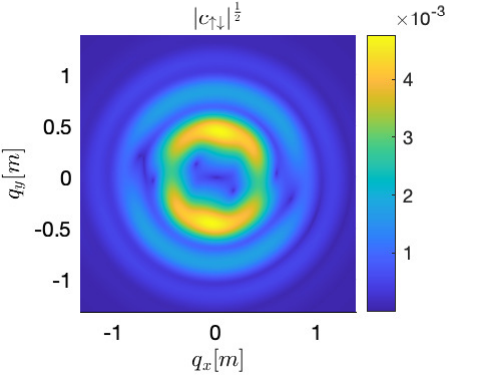}
      \includegraphics[width=0.24\linewidth]{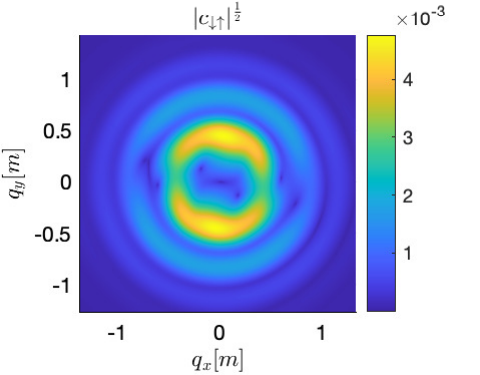}
      \includegraphics[width=0.25\linewidth]{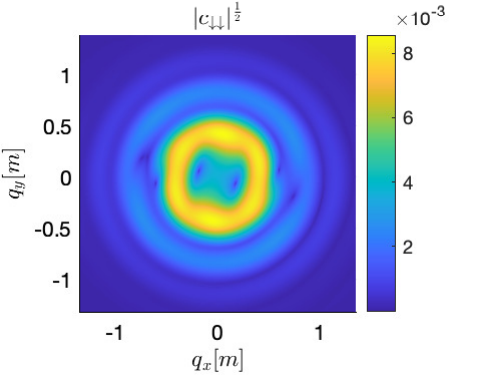}
      \includegraphics[width=0.24\linewidth]{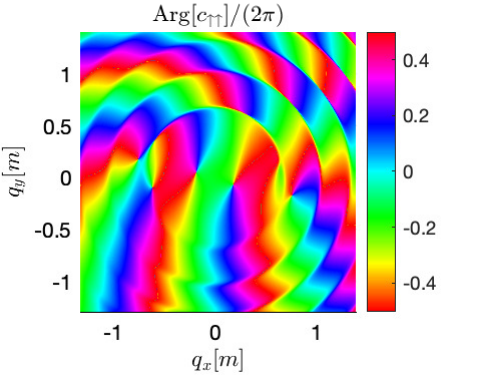}
      \includegraphics[width=0.24\linewidth]{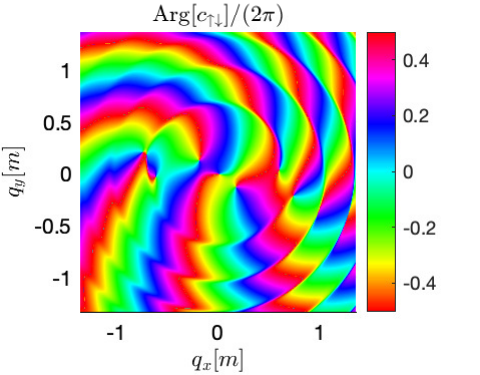}
      \includegraphics[width=0.24\linewidth]{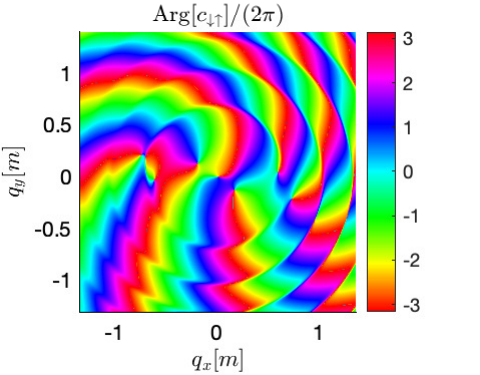}
      \includegraphics[width=0.25\linewidth]{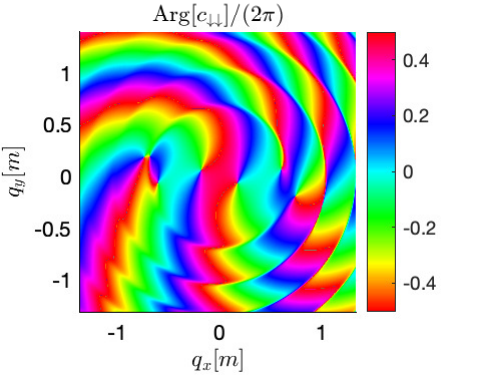}
      \caption{Particle momentum distributions (upper panels) and corresponding phase distributions (lower panels) for electron--positron pair production in the two-color cross field with time-delay parameter $G=0.5$. Other parameters are the same with Fig.~\ref{Fig:Data1}.}\label{Fig:Data3}
 \end{figure*}

 As the time delay is further increased to $G=0.5$ (Fig.~\ref{Fig:Data3}), this incipient vortex structure undergoes a profound topological evolution. The momentum distributions become progressively distorted and asymmetric, with the multi-photon resonance rings exhibiting pronounced azimuthal modulation. Correspondingly, the phase distributions display strongly deformed spiral patterns. The most striking feature of this intermediate regime is the distinct emergence of an organized topological lattice analogous to a von Kármán vortex street. This configuration is a quantum analog of the von Kármán vortex street observed in classical fluid dynamics. In our system, the 'flow' represents the probability current in momentum space, and the 'bluff body' causing the vortex shedding is the temporal gap between the two laser pulses. The vortices self-organize into this staggered array to accommodate the steep phase gradients generated by the interference of time-separated multiphoton absorption pathways. The isolated, symmetric vortex configurations observed at $G=0.2$ are displaced and reorganize into staggered, alternating arrays of vortex–antivortex pairs. Consequently, the associated phase winding becomes tightly compressed along specific angular directions.This structural evolution—from isolated vortex pairs at $G=0.2$ to a fully developed defect lattice at $G=0.5$—reflects the increasing influence of the time delay $G$. Fundamentally, increasing $G$ progressively breaks the temporal parity and residual symmetry of the external driving field, which facilitates a net transfer of orbital angular momentum to the electron-positron pairs. As the temporal separation between the overlapping field components grows, the competing spin-dependent quantum trajectories accumulate significantly different dynamical phases. The coherent superposition of these asymmetric, multi-photon transition amplitudes generates a rapidly rotating phase gradient in momentum space. To accommodate these steep phase gradients while maintaining overall topological charge neutrality, the interference nodes twist into quantized topological defects that naturally distribute into the stable, staggered von Kármán-like lattice. Therefore, the transition from $G=0$ through finite $G$ represents not merely a quantitative change in the interference pattern, but a continuous, spin-dependent dynamic reorganization of the phase topology.

\begin{figure*}[!htb]
      \includegraphics[width=0.24\linewidth]{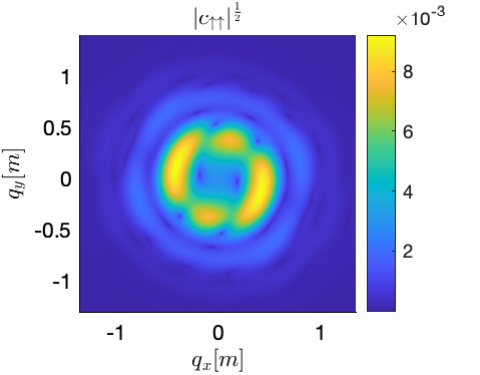}
      \includegraphics[width=0.24\linewidth]{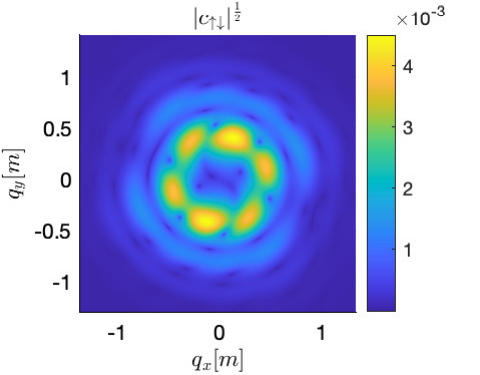}
      \includegraphics[width=0.24\linewidth]{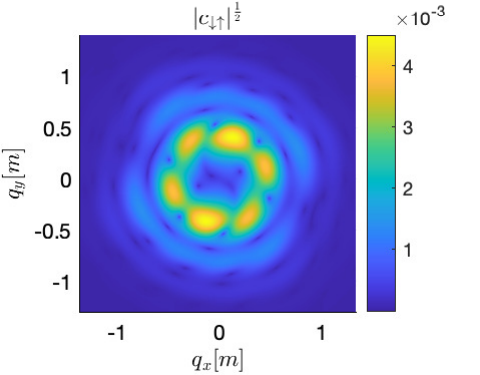}
      \includegraphics[width=0.25\linewidth]{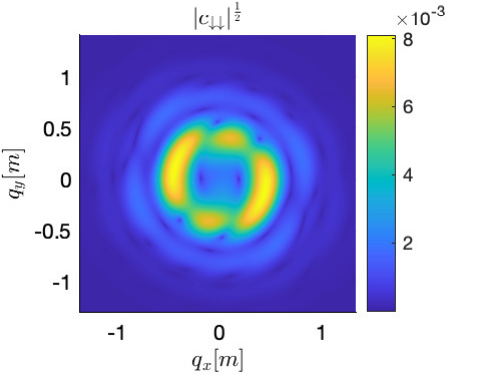}
      \includegraphics[width=0.24\linewidth]{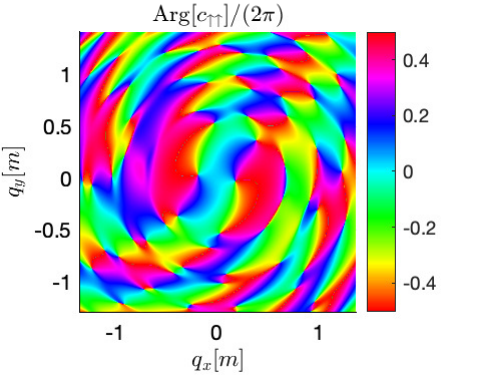}
      \includegraphics[width=0.24\linewidth]{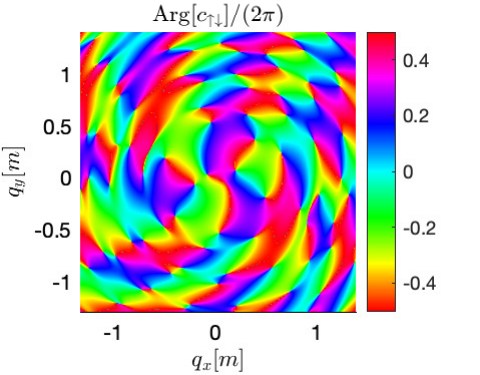}
      \includegraphics[width=0.24\linewidth]{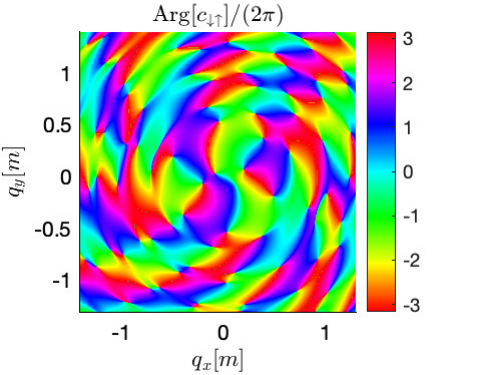}
      \includegraphics[width=0.25\linewidth]{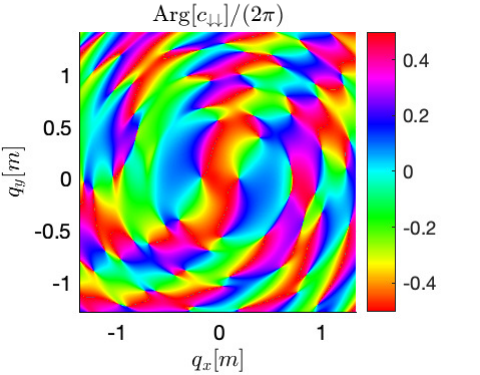}
      \caption{Particle momentum distributions (upper panels) and corresponding phase distributions (lower panels) for electron--positron pair production in the two-color cross field with time-delay parameter $G=1$. Other parameters are the same with Fig.~\ref{Fig:Data1}.}\label{Fig:Data4}
 \end{figure*}

The momentum and phase distributions at a time delay of $G=1$ are presented in Fig.~\ref{Fig:Data4}. In this regime, the system transitions into a fully developed multi-photon resonance state, characterized by the formation of closed concentric energy shells. However, the particle distributions exhibit pronounced azimuthal fragmentation, with multiple localized intensity maxima—a distinct "beaded" morphology—emerging along the ring structures. Compared to the intermediate-$G$ cases, this highly modulated density pattern indicates a substantial enhancement of multi-photon interference, where the precise nodal geometry is dictated by the coherent superposition of competing quantum trajectories.The corresponding phase distributions (lower panels) elucidate the topological origin of this momentum fragmentation and reveal a significant shift in the system's topological behavior. Notably, the distinct, highly organized vortex structures observed at moderate delays begin to degrade. While the phase maps still display a dense proliferation of spiral singularities extending across the momentum plane, the previously clear lattice structures (analogous to von Kármán vortex streets) become less pronounced and less ordered. The periodic vortex cores—points at which the phase is undefined and the wavefunction magnitude must strictly vanish—still map directly onto the interference nodes that segment the momentum rings. However, this degradation of the ordered vortex lattice signals that the transition amplitude is no longer governed by a few dominant quantum pathways. Instead, a multitude of multi-photon channels now contribute with comparable weight, leading to strong phase mixing and an intricate, highly compressed topological state defined by a complex, somewhat chaotic network of vortex–antivortex pairs.Physically, the increased temporal delay $G=1$ further breaks the residual temporal symmetry of the external fields, facilitating a robust transfer of orbital angular momentum to the vacuum excitation. As the number of significant multi-photon interference channels increases, spin-orbit coupling effects become highly pronounced. The relative phases between these numerous spin-dependent transition amplitudes vary so rapidly in momentum space that the clean macroscopic vortex structures are overwhelmed by localized, high-frequency interference fringes. The $G=1$ case thus represents a strongly nonlinear regime in which both the magnitude and the phase topologies of the pair production amplitude are fundamentally governed by dense, multi-channel quantum interference, causing the organized topological arrays to blur into a highly fragmented state.

\begin{figure*}[!htb]
      \includegraphics[width=0.24\linewidth]{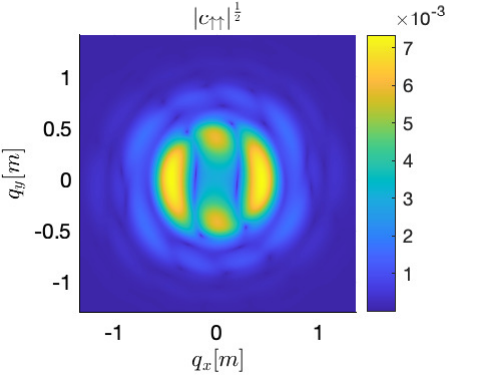}
      \includegraphics[width=0.24\linewidth]{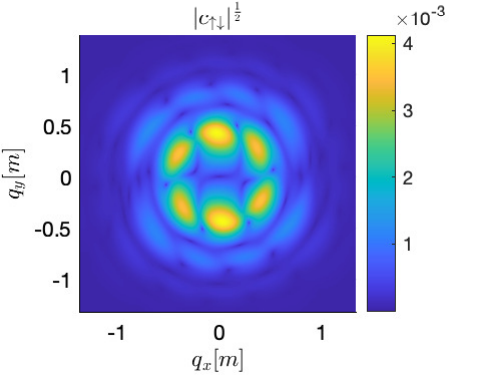}
      \includegraphics[width=0.24\linewidth]{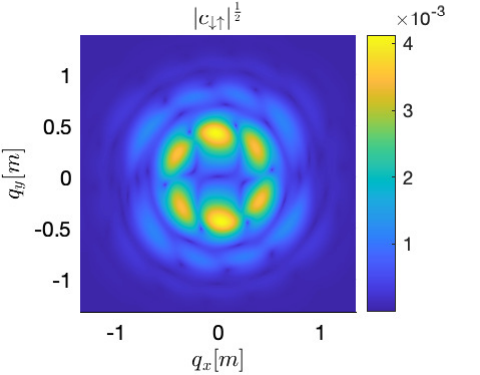}
      \includegraphics[width=0.25\linewidth]{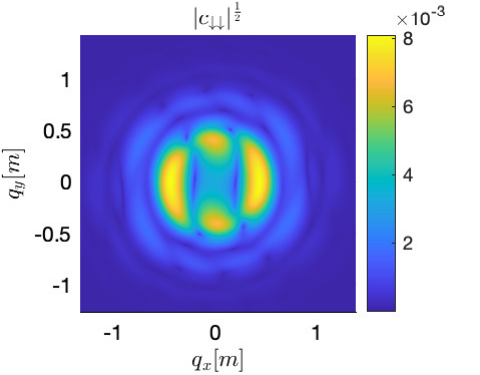}
      \includegraphics[width=0.24\linewidth]{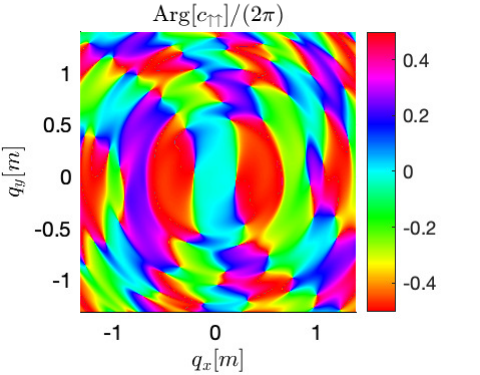}
      \includegraphics[width=0.24\linewidth]{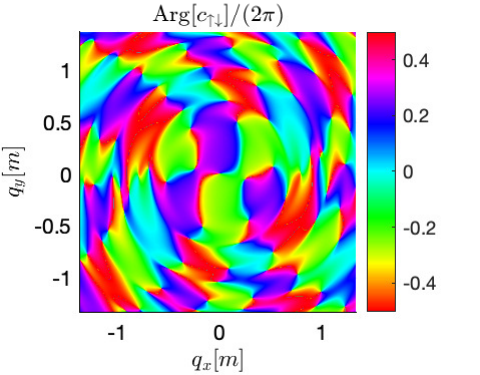}
      \includegraphics[width=0.24\linewidth]{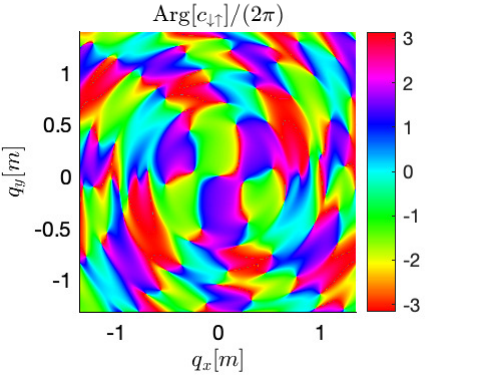}
      \includegraphics[width=0.25\linewidth]{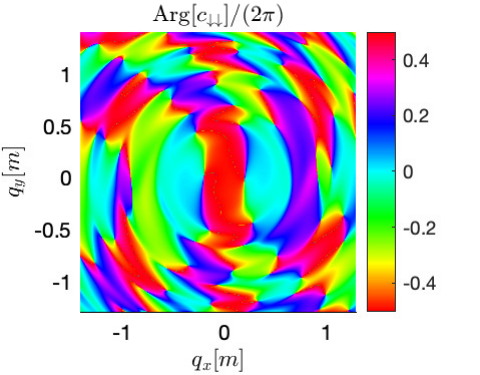}
      \caption{Particle momentum distributions (upper panels) and corresponding phase distributions (lower panels) for electron--positron pair production in the two-color cross field with time-delay parameter $G=2$. Other parameters are the same with Fig.~\ref{Fig:Data1}.}\label{Fig:Data5}
 \end{figure*}
At a large temporal delay of $G=2$, the momentum distributions exhibit fully developed concentric resonance rings corresponding to higher-order multi-photon absorption. The dominant feature in this regime is the severe azimuthal fragmentation of these continuous energy shells into highly localized intensity maxima. Furthermore, the spin state of the created pairs imposes strict geometric constraints on this fragmentation: parallel spin channels ($\uparrow\uparrow$, $\downarrow\downarrow$) exhibit prominent dipole-like splittings, whereas anti-parallel channels ($\uparrow\downarrow$, $\downarrow\uparrow$) resolve into distinct quadrupole structures. This behavior is fundamentally governed by the conservation of total angular momentum $J_z = L_z + S_z$ \cite{Bliokh:2022l}. For parallel spin configurations ($S_z = \pm 1$), the spin triplet-like state carries a portion of the angular momentum, allowing the orbital component $L_z$ to take lower-order values, which results in the observed dipole-like (two-lobed) fragmentation. In contrast, for anti-parallel configurations ($S_z = 0$), the total angular momentum of the absorbed photons must be carried entirely by the orbital motion ($L_z = 2$), naturally producing a quadrupole (four-lobed) nodal structure. This confirms that the spin state acts as a topological filter for the created pairs. Concurrently, the corresponding phase maps reveal a profound topological breakdown. In stark contrast to the organized spiral singularities and lattice structures observed at moderate delays, the macroscopic vortex configurations completely dissolve. The phase topology is instead characterized by remarkably steep, high-frequency oscillatory fringes, forming a highly complex and unstructured landscape devoid of coherent topological defects. This complete suppression of ordered vortex structures is driven by the substantial temporal separation between the overlapping field components, which maximizes temporal symmetry breaking. In this strongly nonlinear regime, pair production is governed by the coherent superposition of an immense number of competing multi-photon trajectories. Because these distinct ionization pathways accumulate significantly different and rapidly varying dynamical phases across momentum space, their interference results in extreme, chaotic phase mixing. This massive multi-channel interference effectively destroys the coherent orbital angular momentum transfer required to sustain stable macroscopic vortices.Nevertheless, the pronounced azimuthal fragmentation of the momentum rings persists as a direct consequence of this highly oscillatory phase landscape. The dense, unstructured phase fringes act as a complex angular grating, where massive destructive interference among the numerous transition amplitudes strictly drives the production probability to zero at localized nodes. Crucially, while macroscopic topological coherence is lost to phase mixing, the system fundamentally conserves total angular momentum. The differing geometric phases accumulated by the spin-flip versus non-spin-flip transitions survive this chaotic interference, continuing to robustly dictate the distinct spin-dependent dipole and quadrupole nodal geometries.

  \section{Summary}\label{sec:4}
  In summary, we have systematically investigated the topological and morphological evolution of spin-resolved multiphoton electron-positron pair creation in time-delayed external electromagnetic fields. By treating the temporal delay $G$ as a continuous tuning parameter, we analyzed the transition of the quantum vacuum excitation from a highly symmetric, interference-suppressed state to a strongly nonlinear multiphoton resonance regime.

Our results reveal that the momentum-space topology is highly sensitive to the temporal overlap of the field components. At $G=0$, the inherent temporal parity of the overlapping fields suppresses topological excitations, resulting in trivial phase domains. As the delay increases ($G=0.2$ to $0.5$), this symmetry is broken, leading to the nucleation of quantized spiral singularities. These singularities self-organize into stable, staggered lattices analogous to von Kármán vortex streets in classical fluid dynamics. In this hydrodynamic-like representation, the probability current in momentum space acts as the flow, while the temporal gap between pulses acts as the symmetry-breaking obstacle.

Furthermore, we demonstrated that the spatial symmetries of the created pairs are fundamentally governed by spin-orbit selection rules and the conservation of total angular momentum $J_z = L_z + S_z$. We found that the spin state acts as a vital geometric filter: parallel spin configurations ($\uparrow\uparrow, \downarrow\downarrow$) consistently enforce a dipole-like (two-lobed) connectivity along the resonance rings, whereas anti-parallel configurations ($\uparrow\downarrow, \downarrow\uparrow$) resolve into distinct quadrupole (four-lobed) structures.

Crucially, while macroscopic vortex coherence dissolves into a chaotic, highly oscillatory phase landscape at large delays ($G \ge 1$) due to massive multi-channel interference, these spin-dependent nodal constraints remain remarkably robust. The persistence of the "beaded" morphology in the resonance rings, even in the absence of ordered phase lattices, confirms that the geometric phase associated with spin-orbit coupling is a fundamental signature of the vacuum excitation. These findings suggest that the topological architecture of momentum space can serve as a high-fidelity diagnostic for the underlying quantum dynamics and the temporal phase structure of ultrastrong laser fields.

  \section{Acknowledgments}\label{sec:8}

Some helpful discussions with A. Orkash is acknowledged. This work was supported by the National Natural Science Foundation of China(NSFC) under Grants No.11974419, No. 12375240, No.12535015 and the Strategic Priority Research Program of Chinese Academy of Sciences under Grants No. XDA25051000 and XDA25010100.

\end{document}